\documentclass[10pt]{article}
\usepackage{amsmath}
\usepackage{amssymb}
\usepackage[Gray,squaren,thinqspace,thinspace]{SIunits}
\usepackage{graphicx}
\usepackage[small,bf]{caption}
\usepackage{a4wide,color}
\usepackage{slashed}
\usepackage[T1,T5]{fontenc}
\usepackage{bbm,url}
\newcommand{\mx}{\ensuremath{\mathsf}}
\DeclareMathOperator{\tr}{tr}

\newcommand{\p}{\partial}

\begin{document}

\title{{\bf Gauge copies in the Landau--DeWitt gauge: a background invariant restriction}}
\author{\textbf{David Dudal}$^{a,b}$\thanks{david.dudal@kuleuven.be}~~\textbf{and}~\textbf{David Vercauteren}$^{c}$\thanks{vercauterendavid@dtu.edu.vn}\\\\
{\small \textnormal{$^{a}$ KU Leuven Campus Kortrijk -- Kulak, Department of
Physics,  }} \\
{\small \textnormal{\phantom{$^{a}$} Etienne Sabbelaan 53 bus 7657, 8500 Kortrijk, Belgium}}\\
{\small \textnormal{$^{b}$ Ghent University, Department of Physics and
Astronomy}}\\
{\small \textnormal{\phantom{$^{b}$}  Krijgslaan 281-S9, 9000 Gent, Belgium}}\\
{\small \textnormal{$^{c}$ Duy T\^an University, Institute of Research and
Development,}}\\{\small \textnormal{\phantom{$^{c}$} P809, 3 Quang Trung, {\fontencoding{T5}\selectfont H\h ai
Ch\^au, \DJ \`a N\~\abreve ng}, Vietnam}} }
\date{}
\maketitle

\begin{abstract}
The Landau background gauge, also known as the Landau--DeWitt gauge, has found renewed interest during the past decade given its usefulness in accessing the confinement-deconfinement transition via the vacuum expectation value of the Polyakov loop, describable via an appropriate background. In this Letter, we revisit this gauge from the viewpoint of it displaying gauge (Gribov) copies. We generalize the Gribov--Zwanziger effective action in a BRST and background invariant way; this action leads to a restriction on the allowed gauge fluctuations, thereby eliminating the infinitesimal background gauge copies. The explicit background invariance of our action is in contrast with earlier attempts to write down and use an effective Gribov--Zwanziger action. It allows to address certain subtleties arising in these earlier works, such as a spontaneous and thus spurious Lorentz symmetry breaking, something which is now averted.
\end{abstract}

\section{Introduction}
A powerful quantization procedure for locally gauge invariant Yang--Mills theories is the background field formalism, in which formalism the gauge field is split in a non-propagating ``classical'' background and a fluctuating quantum part which is integrated over in the path integral procedure. Just as when dealing with an ordinary gauge theory, the quantum gauge fields need to be gauge fixed in the continuum. A particularly useful class of gauges in this context are the background covariant gauges; in these gauges, the background field formalism possesses the important property that, after gauge fixing of and integration over the quantum fields, the eventual (effective) action ought to still be invariant with respect to gauge transformations of the background fields. Useful references are \cite{Abbott:1980hw,Abbott:1981ke}.

Background gauges found a renewed interest during the past decade thanks to their usefulness in probing a typical (non-local) order parameter for the deconfinement transition, the Polyakov loop, whose behaviour can be encoded in a simple specific background, see \cite{Marhauser:2008fz,Braun:2007bx,Reinosa:2014ooa,Reinosa:2015gxn,Reinhardt:2012qe,Reinhardt:2013iia,Cyrol:2017qkll}. Also in the pinch technique combined with Dyson--Schwinger equations, the background field formalism plays a central role \cite{Aguilar:2006gr,Aguilar:2008xm,Binosi:2009qm}. Algebraic aspects of a specific background gauge, the Landau--DeWitt one, including an all order renormalizability proof, were considered in \cite{brstbackground,Ferrari:2000yp,Binosi:2012st}.

Albeit powerful, the (covariant) background gauges are also not free from the famous Gribov ambiguity \cite{Gribov:1977wm} hampering the quantization: multiple gauge equivalent copies of a given quantum gauge field obey the same gauge condition. To deal with this ambiguity, one possibility is to further constrain the space of gauge configurations to be integrated over in the path integral, a procedure proposed by Gribov in \cite{Gribov:1977wm} and worked out by Zwanziger in e.g.~\cite{Zwanziger:1989mf,Zwanziger:1992qr} for the standard Landau gauge. The end point is an effective action ---the Gribov--Zwanziger action--- implementing this restriction. More references can be found in \cite{Vandersickel:2012tz}.

In the presence of backgrounds, seminal work is \cite{Zwanziger:1982na}, based on which a background version of the Gribov--Zwanziger effective action was proposed and used to probe non-perturbative finite temperature dynamics in \cite{Canfora:2015yia,Canfora:2016ngn}.

In this Letter, we revisit in Section 2 the problem of Gribov copies in the Landau--DeWitt gauge and try to \emph{derive} a Gribov--Zwanziger action. Although succeeding in the latter, we identify a major drawback, shared with the \emph{conjectured} action in \cite{Canfora:2015yia,Canfora:2016ngn}: even at zero temperature, a non-zero value of a Lorentz symmetry breaking background is energetically favoured. The problem is traced back to the lack of background gauge invariance and, underlyingly, of BRST invariance of the original Gribov--Zwanziger approach. Motivated by the observation in \cite{brstbackground} that in the background field formalism, BRST invariance at the quantum level is closely linked to background gauge invariance at the classical level, in Section 3 we then go on remedying this problem by constructing a BRST and background invariant version of the Gribov--Zwanziger action, the latter still capable of mitigating the Gribov copy problem but no longer exhibiting the undesirable unphysical features at zero temperature.

\section{Gribov--Zwanziger with a background}
Let us initially work in a general SU($N$) gauge theory, such that the structure constants will be written as $f^{abc}$. Later we will restrict to SU(2) and choose a particular form for the background field.

Our objective is to compute the path integral of Yang--Mills theory in perturbation theory around a given background $\bar A_\mu^a$. We therefore split the total gluon field $a_\mu^a$ as $\bar A_\mu^a + A_\mu^a$, where $A_\mu^a$ are the quantum fluctuations around the classical background field $\bar A_\mu^a$. Instead of the usual Landau gauge $\partial_\mu a^a_\mu=0$ we will choose the Landau background gauge or Landau--DeWitt gauge $\bar{\mathcal D}_\mu^{ab} A_\mu^b = 0$, where $\bar{\mathcal D}_\mu^{ab} = \delta^{ab} \partial_\mu - gf^{abc}\bar A_\mu^c$ is the covariant derivative containing the background field.

We will now give a first way to adapt the Gribov--Zwanziger framework \cite{Gribov:1977wm,Zwanziger:1989mf,Zwanziger:1992qr} to the case with a background, naively following the same steps as led to the original framework without background.

In case of the (transverse) Landau gauge, $\p_\mu A_\mu^a=0$, the Gribov--Zwanziger action arises from the restriction of the domain of integration in the Euclidean functional integral to the so-called Gribov region $\Omega$, which is defined as the set of all gauge  field configurations fulfilling the gauge $\partial_\mu A^{a}_\mu=0$  and for which the (Hermitian) Faddeev--Popov operator is strictly positive. Indeed, requiring positivity of the Faddeev--Popov operator excludes infinitesimal gauge copies, as such copies are connected as $A_\mu^a\to A_\mu^a+ \mathcal{D}_\mu^{ab}\omega^b$ and can both be transverse whenever $-\p_\mu \mathcal{D}_\mu^{ab}\omega^b=0$.

Generalizing to the case at hand, this means that we should restrict to
\begin{equation}
\Omega = \{ A^a_{\mu} \ | \ \bar{\mathcal D}_\mu^{ab} A_\mu^b = 0 \ \& \ {\cal M}^{ac}=-\bar{\mathcal D}_\mu^{ab}(\bar{\mathcal D}_\mu^{bc} - g f^{bcd} A^d_\mu) >0 \} \;. \label{gr}
\end{equation}

The starting point is the (Euclidean) Faddeev--Popov action in the chosen gauge:
\begin{equation} \label{fp}
S_\text{FP} = \int d^{4}x \left( \frac14 F^{a}_{\mu \nu}F^{a}_{\mu\nu} + b^{a}\bar{\mathcal D}_\mu^{ab} A^b_\mu +\bar{c}^{a} \bar{\mathcal D}^{ab}_\mu (\bar{\mathcal D}^{bc}_\mu-gf^{bcd}A^d_\mu)c^c  \right) \;,
\end{equation}
where $({\bar c}^a, c^a)$ stand for the Faddeev--Popov ghosts, $b^a$ is the Lagrange multiplier implementing the Landau gauge, and $F^{a}_{\mu\nu}$ denotes the field strength containing the full gluon field $a_\mu^a = \bar A_\mu^a + A_\mu^a$:
\begin{equation}
F^a_{\mu\nu} = \partial_\mu a^a_\nu - \partial_\nu a^a_\mu + gf^{abc}a^b_\mu a^c_\nu\;. \label{fstr}
\end{equation}
As is generally known, this formalism restricts the path integral to those gluon field configurations obeying the gauge condition. However, this construction still includes many configurations for which the Faddeev--Popov operator is not strictly positive.

In order to impose the second condition, we will construct the no-pole condition \cite{Gribov:1977wm}. Following \cite{Capri:2012wx} we can compute this condition to all orders instead of expanding in the gluon field. We aim to invert the Faddeev--Popov operator $\mathcal M^{ac} = -\bar{\mathcal D}_\mu^{ab} (\bar{\mathcal D}_\mu^{bc} - g f^{bcd} A_\mu^d)$. Let us introduce the operator $\sigma^{ab}$ --- which depends on $\bar A_\mu^a$ and on $A_\mu^a$ --- as
\begin{equation} \label{defsigma}
	(\mathcal M^{-1})^{ae} = \left(\frac1{-\bar{\mathcal D}^2}\right)^{ab} \left(\delta^{be} - g\bar{\mathcal D}_\mu^{bc} f^{cdf} A_\mu^f \left(\frac1{-\bar{\mathcal D}^2}\right)^{de} + \sigma^{be}\right) \;.
\end{equation}
The condition of having a strictly positive Faddeev--Popov operator is equivalent to requiring that the sum of all connected diagrams contributing to the ghost propagator be always finite \cite{Capri:2012wx}, or that
\begin{equation}
	\mathcal G(k) = \langle\mathcal G(k,A)\rangle^\text{conn} = \langle\mathcal M^{-1}\rangle^\text{conn} < \infty \,, \forall k \;.
\end{equation}
Using the expression \eqref{defsigma} and taking account of the facts that the background covariant derivatives $\bar{\mathcal D}_\mu$ do not interact with the taking of the vacuum expectation value (meaning the derivatives can be put in front of the brackets) and that the quantum fields have vanishing vacuum expectation value $\langle A_\mu^a \rangle = 0$, we find that the condition for a strictly positive Faddeev--Popov operator can be written as
\begin{equation}
	\mathcal G^{ac}(k) = \left(\frac1{-\bar{\mathcal D}^2}\right)^{ab} \langle\delta^{bc} + \sigma^{bc}\rangle^\text{conn} = \left(\frac1{-\bar{\mathcal D}^2}\right)^{ab} \left(\frac1{1 - \langle\sigma\rangle^\text{1PI}}\right)^{bc} < \infty \,, \forall k \;.
\end{equation}
In our case, the operator $-\bar{\mathcal D}^2$ will be positive definite (apart from the usual trivial zero), such that the condition we are seeking is equivalent to the no-pole condition that the eigenvalues of $\langle\sigma\rangle^\text{1PI}$ be less than one.

As usual, we will introduce the no-pole condition into the path integral by using the Fourier representation of the Heaviside function:
\begin{equation}
	\int_{-i\infty+\epsilon}^{+i\infty+\epsilon} \frac{d\beta}{2\pi i\beta} e^{\beta - \beta \sigma(0)} \;,
\end{equation}
where the notation $\sigma(0)$ stands for the highest eigenvalues of the $\sigma$ operator. In the case without background, this means the momentum flowing through $\sigma$ is simply set to zero. As a result of this way of introducing the no-pole condition, an extra part $-\beta + \beta \sigma(0)$ will be added to the action. The integration variable $\beta$ will be called the Gribov parameter. The integral over the Gribov parameter is normally done using the steepest descent method, which leads to a gap equation for $\beta$.

Let us now solve \eqref{defsigma} for $\sigma$. After some trivial reordering, we can write that
\begin{equation}
	\sigma^{ab} = (-\bar{\mathcal D}^2)^{ac} (\mathcal M^{-1})^{cb} - \delta^{ab} + g\bar{\mathcal D}_\mu^{ac} f^{cdf} A_\mu^f \left(\frac1{-\bar{\mathcal D}^2}\right)^{db} \;.
\end{equation}
Now replace the Kronecker delta by $-\bar{\mathcal D}^2 \mathcal M \mathcal M^{-1} (-\bar{\mathcal D}^2)^{-1}$:
\begin{multline}
	\sigma^{ab} = (-\bar{\mathcal D}^2)^{ac} (\mathcal M^{-1})^{cd} \left(\delta^{db} - \mathcal M^{de} \left(\frac1{-\bar{\mathcal D}^2}\right)^{eb}\right) + g\bar{\mathcal D}_\mu^{ac} f^{cdf} A_\mu^f \left(\frac1{-\bar{\mathcal D}^2}\right)^{db} \\
	= - g(-\bar{\mathcal D}^2)^{ac} (\mathcal M^{-1})^{cd} \bar{\mathcal D}_\mu^{de} f^{efh} A_\mu^h \left(\frac1{-\bar{\mathcal D}^2}\right)^{fb} + g\bar{\mathcal D}_\mu^{ac} f^{cdf} A_\mu^f \left(\frac1{-\bar{\mathcal D}^2}\right)^{db} \\
	= - g \bigg((-\bar{\mathcal D}^2)^{ac} (\mathcal M^{-1})^{cd} - \delta^{ad}\bigg) \bar{\mathcal D}_\mu^{de} f^{efh} A_\mu^h \left(\frac1{-\bar{\mathcal D}^2}\right)^{fb} \;,
\end{multline}
where, to go the second line, we used the explicit form of $\mathcal M$. Now we replace the Kronecker delta by $\mathcal M (-\bar{\mathcal D}^2)^{-1} (-\bar{\mathcal D}^2) \mathcal M^{-1}$ to get
\begin{multline} \label{uitwerking}	
	\sigma^{ab} = - g \left(\delta^{ad} - \mathcal M^{ac}\left(\frac1{-\bar{\mathcal D}^2}\right)^{cd}\right) (-\bar{\mathcal D}^2)^{de} (\mathcal M^{-1})^{ef} \bar{\mathcal D}_\mu^{fg} f^{ghj} A_\mu^j \left(\frac1{-\bar{\mathcal D}^2}\right)^{hb} \\
	= g^2 \bar{\mathcal D}_\nu^{ac} f^{cdh} A_\nu^h (\mathcal M^{-1})^{de} \bar{\mathcal D}_\mu^{ef} f^{fgj} A_\mu^j \left(\frac1{-\bar{\mathcal D}^2}\right)^{gb} \\
	= g^2 \bar{\mathcal D}_\nu^{ac} f^{cdh} A_\nu^h (\mathcal M^{-1})^{de} f^{efj} A_\mu^j \bar{\mathcal D}_\mu^{fg} \left(\frac1{-\bar{\mathcal D}^2}\right)^{gb} \;,
\end{multline}
where we again used the explicit form of $\mathcal M$ when going to the second line, and where we furthemore used that $\bar{\mathcal D}_\mu^{ab} f^{bcd} A_\mu^d = f^{abd} A_\mu^d \bar{\mathcal D}_\mu^{bc}$ if the quantum gluon field $A_\mu^a$ obeys the Landau background gauge condition $\bar{\mathcal D}_\mu^{ab} A_\mu^b$.\footnote{We will always use backgrounds that obey the Landau gauge condition $\partial_\mu \bar A_\mu^a = 0$ themselves.}

Finally, the no-pole condition will be imposed on
\begin{equation} \label{nopolegeneral}
	\langle\sigma^{ab}\rangle^\text{1PI} = g^2 \bar{\mathcal D}_\nu^{ac} \left\langle f^{cdh} A_\nu^h \left(\frac1{\mathcal M}\right)^{de} f^{efj} A_\mu^j\right\rangle^\text{1PI} \bar{\mathcal D}_\mu^{fg} \left(\frac1{-\bar{\mathcal D}^2}\right)^{gb} \;.
\end{equation}

From now on we will restrict ourselves to the SU(2) gauge group. We will restrict ourselves here to a background of the type
\begin{equation}\label{dd0}
\bar A_\mu^a = \mathcal A \delta^{a3} \delta_{\mu0}.
 \end{equation}
given its practical relevance in relation to the vacuum expectation value of the Polyakov loop \cite{ Marhauser:2008fz,Braun:2007bx,Reinosa:2014ooa,Reinosa:2015gxn,Reinhardt:2012qe,Reinhardt:2013iia,Cyrol:2017qkll,Canfora:2015yia}.

This background breaks global colour rotation invariance, but there is still invariance in the $(1,2)$ colour plane. Under this kind of rotations, the components of the ghost propagator transform as ($i$ and $j$ can take the values $1$ or $2$): $\mathcal G^{33}$ as a scalar, $\mathcal G^{3i}$ and $\mathcal G^{i3}$ as a vector, and $G^{ij}$ as a tensor. This means that $\mathcal G^{3i} = \mathcal G^{i3} = 0$, and that $\mathcal G^{ij}$ must be a constant matrix $\delta^{ij} \mathcal G$. As a result, the no-pole condition leads to two conditions: one on $\mathcal G^{33}$ and one on $\tfrac12 \mathcal G^{ii}$.

If we now introduce the isospin-shifted plane waves, using the isospin eigenstates defined in \eqref{isospin},
\begin{equation} \label{eigenstates}
	|p_\mu,0\rangle = e^{ip_\mu x_\mu} \mx v_3 \;, \qquad |p_\mu,s\rangle = e^{ip_\mu x_\mu + igs\mathcal Ax_0} \mx v_s
\end{equation}
for $s=\pm$, we find that $\mathcal D_\mu |p_\nu,s\rangle = ip_\mu |p_\nu,s\rangle$ for $s=\pm,0$.

\subsection{The $3$ sector}
In the $3$ sector the no-pole condition becomes
\begin{equation}
	g^2 \partial_\nu \left\langle \epsilon^{3ac} A_\nu^c \left(\frac1{\mathcal M}\right)^{ab} \epsilon^{b3d} A_\mu^d\right\rangle^\text{1PI} \partial_\mu \frac1{-\partial^2} < 1 \;.
\end{equation}
This can be written using the $|p_\mu,0\rangle$ states defined in \eqref{eigenstates}, yielding
\begin{equation}
	g^2 \frac{p_\mu p_\nu}{p^2} \sum_{s,t=\pm} s t \left\langle A_\nu^s \left(\mx v_s^\dagger \frac1{\mathcal M} \mx v_t\right) A_\mu^{-t} \right\rangle^\text{1PI} < 1 \;,
\end{equation}
where we made use of the fact that momentum is conserved. In the $3$ sector, the lowest eigenvalue of the Faddeev--Popov operator in the absence of quantum fluctuations is for the (non-shifted) momentum equal to zero. Like is usually done, we can thus take the limit $p \to 0$ to find:
\begin{equation}
	\frac{g^2}{Vd} \sum_{s,t=\pm} s t \left\langle p=0 \left| A_\mu^s \left(\mx v_s^\dagger \frac1{\mathcal M} \mx v_t\right) A_\mu^{-t} \right| p=0 \right\rangle^\text{1PI} < 1 \;.
\end{equation}

When imposing the no-pole condition through the Fourier representation of the Heaviside function, a term
\begin{equation}
	\frac{g^2\beta_3}{Vd} \sum_{s,t=\pm} s t \int d^dx d^dy \; A_\mu^s(x) \left(\mx v_s^\dagger \frac1{\mathcal M} \mx v_t\right)(x-y) A_\mu^{-t}(y)
\end{equation}
will have to be added to the action. If we want to compute the quadratic order in the quantum fields, we can replace the Faddeev--Popov operator by its lowest order form $-\bar{\mathcal D}^2$, for which we have that $-\bar{\mathcal D}^2 e^{ip_\mu x_\mu} \mx v_s = ((p_0-gs\mathcal A)^2 + \vec p^2)e^{ip_\mu x_\mu} \mx v_s$, with $\vec p$ the vector formed by the $p_\mu$ with $\mu = 1\cdots (d-1)$. As a result, we find that the extra term at quadratic order is equal to
\begin{equation}
	\frac{g^2\beta_3}{Vd} \sum_{s=\pm} \int \frac{d^dp}{(2\pi)^d} \tilde A_\mu^{-s}(-p) \frac1{(p_0+gs\mathcal A)^2+\vec p^2} \tilde A_\mu^s(p) \;,
\end{equation}
where the Fourier transforms are now taken with \emph{unshifted} momentum eigenstates $e^{ip_\mu x_\mu}$.

\subsection{The $1,2$ sector}
For the $1,2$ sector, we work analogously. As the $\sigma$ operator is expected to be a constant isospin matrix in this sector, we can consider half the trace. We thus have the condition
\begin{equation}
	\frac{g^2}2 \sum_{a=1,2} \bar{\mathcal D}_\nu^{ab} \left\langle \epsilon^{bcg} A_\nu^g \left(\frac1{\mathcal M}\right)^{cd} \epsilon^{deh} A_\mu^h\right\rangle^\text{1PI} \bar{\mathcal D}_\mu^{ef} \left(\frac1{-\bar{\mathcal D}^2}\right)^{fa} < 1 \;.
\end{equation}
In order to sum over $a$, we sandwich the above summand between the states $|p_\mu,s\rangle$ defined in \eqref{eigenstates} and sum over $s$. This yields
\begin{equation}
	- \frac{g^2}4 \frac{p_\mu p_\nu}{p^2} \sum_{s=\pm} \int d^dx d^dy \left\langle (\epsilon^{1ac}-is\epsilon^{2ac}) A_\nu^c \left(\frac1{\mathcal M}\right)^{ab} (\epsilon^{b1d}+is\epsilon^{b2d}) A_\mu^d \right\rangle^\text{1PI} < 1 \;.
\end{equation}
In this case, the lowest eigenvalue of the Faddeev--Popov operator is reached for the \emph{shifted} momentum $p_\mu=0$, such that we find for the highest eigenvalue of $\sigma$ in this sector:
\begin{equation}
	- \frac{g^2}{4Vd} \sum_{s=\pm} \int d^dx d^dy \left\langle (\epsilon^{1ac}-is\epsilon^{2ac}) A_\mu^c(x) \left(\frac1{\mathcal M}\right)^{ab}(x-y) (\epsilon^{b1d}+is\epsilon^{b2d}) A_\mu^d(y) \right\rangle^\text{1PI} < 1 \;.
\end{equation}
Using the isospin eigenstates \eqref{isospingluon}, we find for the extra term that has to be added to the Lagrangian
\begin{multline}
	\frac{g^2\beta_{12}}{2Vd} \sum_{s=\pm} \int d^dx d^dy \left( A_\mu^3(x) \left(\mx v_s^\dagger \frac1{\mathcal M} \mx v_s\right)(x-y) A_\mu^3(y) + A_\mu^{-s}(x) \left(\mx v_3^\dagger \frac1{\mathcal M} \mx v_3\right)(x-y) A_\mu^s(y) \right. \\ \left. - A_\mu^3(x) \left(\mx v_s^\dagger \frac1{\mathcal M} \mx v_3\right)(x-y) A_\mu^s(y) - A_\mu^{-s}(x) \left(\mx v_3^\dagger \frac1{\mathcal M} \mx v_s\right)(x-y) A_\mu^3(y) \right) \;.
\end{multline}

If we now again restrict ourselves to quadratic order in the quantum fields, the last two terms vanish because the Faddeev--Popov operator at leading order does not contain any terms mixing the $1,2$ and the $3$ sectors. Proceeding like in the previous case we find
\begin{equation}
	\frac{g^2\beta_{12}}{2Vd} \int \frac{d^dp}{(2\pi)^d} \tilde A_\mu^3(-p) \left( \sum_{s=\pm} \frac1{(p_0+gs\mathcal A)^2+\vec p^2} \right) \tilde A_\mu^3(p) + \frac{g^2\beta_{1,2}}{2Vd} \sum_{s=\pm} \int \frac{d^dp}{(2\pi)^d} \tilde A_\mu^{-s}(-p) \frac1{p^2} \tilde A_\mu^s(p) \;.
\end{equation}

\subsection{Summarizing}
Now introduce the $\lambda$ Gribov parameters as $\lambda_3^4 = 2g^2\beta_3/Vd$ and $\lambda_{12}^4 = g^2\beta_{12}/Vd$. In conclusion, the inverse gluon propagator in the $3$ sector ends up being
\begin{subequations} \label {props} \begin{equation}
	p^2 \left(\delta_{\mu\nu} - \left(1-\frac1\xi\right)\frac{p_\mu p_\nu}{p^2}\right)  + \delta_{\mu\nu} \lambda_{12}^4 \left(\frac1{(p_0+g\mathcal A)^2+\vec p^2} + \frac1{(p_0-g\mathcal A)^2+\vec p^2}\right)
\end{equation}
and the one in the sector with spin $s$ (with $s=\pm$)
\begin{multline}
	((p_0 - gs\mathcal A)^2 + \vec p^2) \left(\delta_{\mu\nu} - \left(1-\frac1\xi\right)\frac{(p_\mu-gs\mathcal A\delta_{\mu0})(p_\nu-gs\mathcal A\delta_{\nu0})}{(p_0-gs\mathcal A)^2+\vec p^2}\right) \\ + \delta_{\mu\nu} \left(\frac{\lambda_3^4}{(p_0 + sg\mathcal A)^2+\vec p^2} + \frac{\lambda_{12}^4}{p^2}\right) \;.
\end{multline} \end{subequations}
One sees that the Gribov formalism adds terms of the shape $1/((p_0+sg\mathcal A)^2+\vec p^2)$ with $s=\pm,0$ to the inverse propagators, but the values of $s$ appearing in the additions are always different from the values appearing in the non-Gribov parts of the inverse propgator. This is of course related to the fact that the quantum gluon field always appears contracted with structure constants in the no-pole condition \eqref{nopolegeneral}.

\subsection{Effective action}
If we want to investigate which value of the background is realized, we need to minimize the effective action. The one-loop correction is given by half the traced logarithm of the propagators in \eqref{props}. (The tree level contribution does not depend on the gluon background, which has zero field strength, and the ghost contributions are also trivial as the presence of the background only shifts the momenta.)

In order to get an idea of what will happen without getting bogged down in technical computations, we will simply probe the effective action around the point $\mathcal A=0$ in the direction of varying $\mathcal A$, i.e.~we will expand in the parameter $\mathcal A$ in the propagators \eqref{props} while keeping the Gribov parameters equal to their $\mathcal A=0$ values: $\lambda_3^4 = \lambda_{12}^4 = \frac12 \lambda^4$.

After some tedious but straightforward algebra, we find for the effective action $\mathcal E$
\begin{multline}
	\mathcal E = - \frac{3d}{4g^2} \lambda^4 + \frac32 (d-1) \tr\log\left(p^2+\frac{\lambda^4}{p^2}\right) \\ + (d-1)g^2\mathcal A^2 \tr\left(\frac{p^4-\lambda^4}{p^2(p^4+\lambda^4)} - \frac{2p_0^2(4p^8-4p^4\lambda^4-7\lambda^8)}{p^4(p^4+\lambda^4)^2}\right) + \mathcal O(\mathcal A^4) \;.
\end{multline}
Let us simply consider the coefficient of $\mathcal A^2$ in the above expression. In the second term within the trace, we can --- for symmetry reasons --- replace $p_0^2$ by $p^2/d$ with $d$ the number of spacetime dimensions. Next, in order to be able to use dimensional regularization, we split the resulting integrand as
\begin{multline}
	\frac{(d-1)g^2}{2d} \int \frac{d^dp}{(2\pi)^d} \left( \frac{2(d-2)p^8 + 4p^4\lambda^4 + (7-2d)\lambda^8}{p^2(p^4+\lambda^4)^2} - \frac{2(d-2)p^2}{p^4+\lambda^4} \right) \\ + \frac{(d-1)g^2}{2d} \int \frac{d^dp}{(2\pi)^d} \frac{2(d-2)p^2}{p^4+\lambda^4} \;.
\end{multline}
The first integral now converges for $2<d<6$ while the second one converges for $-2<d<2$. Evaluating the integrals and expanding around $d=4$ gives only a finite contribution
\begin{equation}
	-\frac{g^2\lambda^2}{(4\pi)^2} \frac{27\pi}{32}
\end{equation}
for the coefficient of $\mathcal A^2$.

On can interpret this result as follows. In the physical vacuum, $\mathcal A$ will change its value so as to minimize the effective action. If we start from the value $\mathcal A=0$, we can see that the effective action under changes of $\mathcal A$ at that point is concave (the coefficient of $\mathcal A^2$ in the expansion is negative), such that any change in $\mathcal A$ (without changing the Gribov parameters) will lead to a lower value for the effective action. Now of course the Gribov parameters will change as well under the influence of the new value of $\mathcal A$, but we know that the physical solution with $\lambda\not=0$ is a maximum of the effective action.\footnote{The value of the Gribov parameter is determined by the steepest descent method, not by minimization of the effective action.} This means that any change in the values of the Gribov parameters near $\mathcal A=0$ will also lower the value of the effective action. In conclusion, a change away from $\mathcal A=0$ will always lower the effective action, such that $\mathcal A=0$ cannot be the physical vacuum.

In order to figure out what the final value of $\mathcal A$ will be, one would need to compute the full effective action. A quick numerical computation shows that it is unbounded from below for variation of $\mathcal A$ keeping the Gribov parameters at their $\mathcal A=0$ value, which suggests there may be no finite solution. But even if giving the Gribov parameters their physical ($\mathcal A$ dependent) values were to lead to a solution at finite $\mathcal A$, this would still spell disaster to the theory as Lorentz symmetry would be spontaneously broken.

\section{Lack of background gauge invariance and a symmetric approach to resolve this}
As shown in the previous section, it turns out that a naive approach to Gribov--Zwanziger in the presence of a background leads to trouble. The solution can be found by using a more symmetric approach.

The ordinary approach to background fields in gauge theory without taking the presence of the Gribov horizon into account, i.e.~based on the action \eqref{fp}, is symmetric under both the usual BRST transformations \cite{brstbackground} \emph{and} under background gauge transformations, where the background transforms like a gauge field while the other fields (the quantum part of the gluon field and the ghosts) transform in the adjoint representation:\footnote{To be more precise: each colour index is associated to an adjoint transformation; the eventual Gribov--Zwanziger action contains auxiliary fields with a double colour index.}
\begin{equation}
	\delta \bar A_\mu^a = \bar{\mathcal D}_\mu^{ab} \omega^b \;, \quad \delta A_\mu^a = gf^{abc} A_\mu^b \omega^c \;\,, \ldots
\end{equation}
Due to this background gauge invariance, the effective action should only depend on gauge invariant combinations of the background, such as its field strength. In the naive approach outlined in the previous section, this is obviously not the case, as a constant $\bar A_\mu^a$ is gauge equivalent with zero, while the effective action constructed in the previous section does depend on the background, leading to the physical problem outlined.

It is, however, not at all straightforward to use the above-mentioned symmetries in combination with the Gribov--Zwanziger formalism. In fact, in \cite{Canfora:2015yia,Canfora:2016ngn} a trial Gribov--Zwanziger action was employed in presence of a background, inspired by \cite{Zwanziger:1982na}. However, a similar observation holds true at $T=0$ when properly\footnote{It should be noted that the $\epsilon$-structure constants coupling the gauge field to the auxiliary fields were (erroneously) neglected in \cite{Canfora:2015yia,Canfora:2016ngn}.} using that background Gribov--Zwanziger action: the ground state would favour a nonzero $\mathcal{A}$.

Only recently has a BRST symmetry consistent with the Gribov construction been found \cite{Capri:2015ixa}. In order to construct a more symmetric formalism including both a background and the restriction to the Gribov horizon, we will take \cite{Capri:2015ixa,Capri:2016aqq,Capri:2017bfd} as our starting point.

\subsection{BRST symmetric formulation}
As the starting point we take the Lagrangian proposed in \cite{Capri:2017bfd}:
\begin{align} \label{symaction}
	S_\text{BRST} &= \int d^4x \left( \frac14 F^a_{\mu \nu}F^a_{\mu\nu} + b^a\partial_\mu a^a_\mu +\bar c^a \partial_\mu (\delta^{ab} \partial_\mu-gf^{abc}a^c_\mu)c^b \right) \nonumber\\
	&+ \int d^4x \left( \bar\phi_\mu^{ac} \partial_\nu (\mathcal D^h)_\nu^{ab} \phi_\mu^{bc} - \bar\omega_\mu^{ac} \partial_\nu (\mathcal D^h)_\nu^{ab} \omega_\mu^{bc} + \gamma^2 f^{abc} (a^h)_\mu^a (\bar\phi_\mu^{bc} + \phi_\mu^{bc})-d(N^2-1)\gamma^4\right)\nonumber\\
&+\int d^4x \left(\bar\eta^a \partial_\nu (\mathcal D^h)_\mu^{ab} \eta^b + \tau^a \p_\mu (a^h)_\mu^a\right) \;.
\end{align}
In this expression, $a_\mu^a$ is still the full gluon field while $a^h$ is a transversal projection of the gluon field, and $\mathcal D^h $ is the covariant derivative using this $a^h$ field:
\begin{equation}
	(\mathcal D^h)_\mu^{ab} = \delta^{ab} \partial_\mu - gf^{abc} (a^h)_\mu^c \;.
\end{equation}
More precisely, $a^h$ is obtained as the configuration corresponding to the minimum of $\int d^4x \; a_\mu^a a_\mu^a$ along the gauge orbit. As discussed in e.g.~\cite{Capri:2015ixa,Lavelle:1995ty}, $a^h$ is not only transversal, but even gauge invariant. As such, it is not difficult to verify the BRST invariance of \eqref{symaction} under the (nilpotent) variation
\begin{equation} \label{brst} \begin{gathered}
  sa_\mu^a=-\mathcal{D}_\mu^{ab}c^b\;,\quad sc^a=\frac{g}{2}f^{abc}c^b c^c\;,\quad s\bar c^a= b^a\;,\quad sb^a=0\;, \\
  s \phi_\mu^{ac}=  s\bar\phi_\mu^{ac}=s\bar\omega_\mu^{ac}=s\phi_\mu^{ac}=s\bar\eta^a=s\eta^a=s \tau^a=0 \;.
\end{gathered} \end{equation}
To be more precise, the stationarity condition applied to $\int d^{4}x \; a_\mu^a a_\mu^a$ leads to $\p a^h=0$, while the minimization condition requires the second variation to be positive, which turns out to be $-\partial_\mu\mathcal D^h_\mu>0$. The latter constraint is enforced at the level of the action via the Gribov--Zwanziger construction, which precisely leads to the expression \eqref{symaction}. The transversal character of $a^h$ is enforced via the Lagrange multiplier field $\tau$, while the $(\bar\eta,\eta)$ are extra Grassmann fields related to the Jacobian accompanying the constraint $\p a^h=0$, see footnote~5 in \cite{Capri:2017bfd}. It is tacitly assumed here that $\p a^h=0$ is, per configuration $a$, always solved for in an iterative way, i.e.~in a formal power series in the coupling $g$ (or powers of the field) around $a$, just as presented in Appendix A of \cite{Capri:2015ixa} for example. This is also the setting in which the renormalizability of the formulation \eqref{symaction} was proven \cite{Capri:2017bfd}. This entails we do not (and cannot) bother about the convergence properties of this series representation of the minimum (considerations of which go beyond the essentially perturbative approach used in this paper). It also means we are considering per configuration $a$ the local minimum of the functional that is connected to it in the just described perturbative sense. The associated positive second variation implies that there are no other, infinitesimally connected, gauge equivalent other solutions to the minimization problem, as this is nothing else than the crux of the Gribov restriction. About possible other non-trivial, globally connected, solutions we have nothing to say; this is a difficult topological question of which is unknown how to treat it in practice \cite{vanBaal:1991zw}.

The Gribov parameter $\gamma^2$ is determined via its gap equation, $\frac{\p \mathcal{E}}{\p \gamma^2}=0$ with $\mathcal{E}$ the (vacuum) effective action. Upon integrating over the auxiliary fields $(\bar\phi,\phi,\bar\omega,\omega)$, this gap equation is equivalent to setting (where eventually $d\to4$ is understood, upon properly dimensionally regularizing and renormalizing the theory)
\begin{equation}
\mathcal{H}(a^h)=dV(N^2-1) \;, \qquad \mathcal{H}(a^h)= g^2\int d^d x d^d y f^{abc}(a^h)_\mu^b(x) \left[-\p_\nu \mathcal{D}_\nu^h\right]^{-1}_{ad} f^{dec}(a^h)_\mu^e(y)\;,
\end{equation}
the BRST invariant version of the original Zwanziger horizon condition. For more details, we refer to \cite{Zwanziger:1989mf,Zwanziger:1992qr,Capri:2012wx,Capri:2015ixa,Capri:2016aqq,Capri:2017bfd}. It ensures the path integral restriction to a region $\Omega$ where $-\p_\mu \mathcal{D}_\mu^h$ has no zero modes, equivalent to the standard Faddeev--Popov operator $-\p_\mu \mathcal{D}_\mu$ having none in the Landau gauge, as $a^h=a$ when on-shell $\p a=0$.

If we want to introduce a background through the usual split $a_\mu^a = \bar A_\mu^a + A_\mu^a$ in the action \eqref{symaction}, the first line of the action must be turned into what we wrote in equation \eqref{fp}. The fate of the second line is more subtle, however. We propose to replace it with
\begin{align} \label{symactionbis}
S_\text{BRST} &= \int d^4x \left(  \frac14 F^{a}_{\mu \nu}F^{a}_{\mu\nu} + b^{a}\bar{\mathcal D}_\mu^{ab} A^b_\mu +\bar{c}^{a} \bar{\mathcal D}^{ab}_\mu (\bar{\mathcal D}^{bc}_\mu-gf^{bcd}A^d_\mu)c^c  \right)\nonumber \\
	&+ \int d^4x \left( \bar\phi_\mu^{ac} \partial_\nu (\mathcal D^h)_\nu^{ab} \phi_\mu^{bc} - \bar\omega_\mu^{ac} \partial_\nu (\mathcal D^h)_\nu^{ab} \omega_\mu^{bc} + \gamma^2 f^{abc} \left[(a^h)_\mu^a -(\bar A^h)_\mu^a\right](\bar\phi_\mu^{bc} + \phi_\mu^{bc}) \right)\nonumber\\
&+\int d^4x \left(-d(N^2-1)\gamma^4+\bar\eta^a \partial_\mu (\mathcal D^h)_\mu^{ab} \eta^b +\tau^a \p_\mu (a^h)_\mu^a\right) \;.
\end{align}
Notice that when coupling the gauge transformed gauge field $a^h$ to the localizing auxiliary fields $(\bar\phi,\phi)$, we have now used $a^h - \bar A^h$ instead. This is because we are only interested in imposing the Gribov condition on the quantum fields, which are the fields we integrate over. This way the series of $a^h - \bar A^h$ still starts with a term of first order in the quantum gauge field with the foregoing action naturally reducing to \eqref{symaction} in case the background is zero modulo a local gauge transformation.

Let us now show why the action \eqref{symactionbis} has the desired property of eliminating infinitesimally equivalent background gauge copies. First of all, the explicit expression of $\mathcal{H}(\bar A^h-a^h)$ and the ensuing implementation of the horizon condition and localization in terms of the action \eqref{symactionbis} is completely similar to the derivation worked out in \cite{Capri:2012wx}. Secondly, by imposing $-\p \mathcal{D}^h\equiv -\partial_\mu\mathcal D^h_\mu >0$, we are actually excluding a large set of Gribov copies related to the zero modes of the Faddeev-Popov operator $-\bar{\mathcal{D}}(-\bar{\mathcal{D}}-A)$ introduced in \eqref{gr}, the original premise of the whole construction. We will use a shorthand notation from here on to avoid clutter of colour and Lorentz indices. We again restrict ourselves to backgrounds of the type \eqref{dd0}.  As shown below, see also \cite{Capri:2015ixa}, $a^h$ is subject to $\p a^h=0$. Setting $a=a^h+\tau$, we get $\p a = \p \tau$, or for our specific background, $\p\tau=\p A$ and thus $\p\tau=\bar{\mathcal{D}}A+\bar A A=\bar A A$ because of the gauge condition. Clearly, $\tau =\mathcal{O}(\bar{A})$. Now, let $\xi$ be a zero mode of $-\bar{\mathcal{D}}(-\bar{\mathcal{D}}-A)$, then equivalently we can write
\begin{eqnarray}\label{dd1}
  \left[-\bar{\mathcal{D}}(-\bar{\mathcal{D}}-A)\right]\xi=0 &\Leftrightarrow& -\p \mathcal{D}\xi=\bar A\mathcal{D}\xi \Leftrightarrow -\p \mathcal{D}^h\xi=-\bar A\mathcal{D}\xi+\p(\tau \xi)\nonumber\\
  &\Leftrightarrow& \xi = -\left[\p\mathcal{D}^h\right]^{-1}\left[-\bar A\mathcal{D}\xi+\p(\tau \xi)\right]
\end{eqnarray}
In the last line, we used the assumed positivity and thus invertibility of $-\p \mathcal D^h$. Writing $\xi=\sum_{n=0}^{\infty}\bar A^n\xi_n[A]$, \eqref{dd1} together with the fact $\tau=\mathcal{O}(\bar A)$, iteratively leads to $\xi_n=0$, viz.~$\xi=0$. Essentially, we have thus shown that the background Faddeev-Popov operator has no zero modes that are expressable as a Taylor series in the constant background field, i.e.~that are continuous deformations around the zero background (standard Landau gauge). Notice that we are actually considering a set of gauges that are simple continuous deformations around the Landau gauge, the deformation parameterized by the background parameter $\mathcal{A}$.

An interesting byproduct is that with the following construction, we will only need a single Gribov parameter $\lambda^4$, in contradistinction with the two parameters that had to be introduced in the previous section.

\subsection{Determination of $a^h$ for an explicit choice of the background}
In order to write down an expression for the BRST invariant gauge field $a^h$, we need to find the gauge transform that brings $a_\mu^a$ from a certain given value to the value that minimizes the integral of its square
\begin{equation} \label{intasquared}
	\int d^4x \; a_\mu^a a_\mu^a \;.
\end{equation}
This means we write
\begin{equation}
	a^h_\mu = h (\bar A_\mu + A_\mu) h^\dagger + \frac ig h\partial_\mu h^\dagger
\end{equation}
with $A_\mu = A_\mu^a \sigma^a/2$ etc. the gauge fields in matrix form and $h$ an SU(2) gauge transform matrix that needs to be determined. We can write $h$ using an expansion in the quantum gauge field:
\begin{equation}
	h = e^{\frac i2 \varphi^a \sigma^a} \;, \quad \varphi^a = \varphi_0^a + \varphi_1^a + \cdots
\end{equation}
where $\varphi_0^a$ is of zeroth order in the quantum gauge field, $\varphi_1^a$ is of first order, and so on.

Before we impose the minimization, we work out some of the quantities involved. We can rewrite the matrix $h$ as
\begin{equation}
	h = \mathbbm1 \cos\frac\varphi2 + i \hat\varphi^a \sigma^a \sin\frac\varphi2 \;,
\end{equation}
where $\varphi$ is the norm of the vector $\varphi^a$ and $\hat\varphi^a = \varphi^a/\varphi$ is the unit vector pointing in the direction of $\varphi^a$. Next, some straighforward algebra leads to the gauge transform
\begin{equation} \begin{aligned}
	h a_\mu h^\dagger + \frac ig h\partial_\mu h^\dagger = & \left( a_\mu^a \cos\varphi - \epsilon^{abc} \hat\varphi^b a_\mu^c \sin\varphi + \hat\varphi^b a_\mu^b \hat\varphi^a (1-\cos\varphi) \phantom{\frac1g} \right. \\
	& \left. + \frac1g \partial_\mu\varphi^a + \frac1g \partial_\mu\hat\varphi^a (\sin\varphi-\varphi) + \frac1g \epsilon^{abc} \hat\varphi^b \partial_\mu\hat\varphi^c (\cos\varphi-1) \right) \frac{\sigma^a}2 \;.
\end{aligned} \end{equation}
This means that $(a^h)_\mu^a$ is the expression between the brackets.

Now let us consider $a^h$ at zeroth order in the quantum fields. It is obvious that, at this order, the expression in \eqref{intasquared} will be minimized if the gauge transform $h$ is the one that brings $\bar A_\mu^a$ to zero, or
\begin{equation} \label{phinaught}
	\varphi_0^a = -g \mathcal A x_0 \delta^{a3} \;.
\end{equation}
Going to first order in the quantum fields, we find after some more tedious but uneventful algebra that
\begin{multline}
	(a^h)_\mu^a = A_\mu^a \cos g\mathcal Ax_0 - \epsilon^{ab3} A_\mu^b \sin g\mathcal Ax_0 + A_\mu^3 \delta^{a3} (1-\cos g\mathcal Ax_0) \\
	+ \frac{\delta_{\mu0}}{gx_0} \left( \delta^{ab} \left(\cos g\mathcal Ax_0 - \frac{\sin g\mathcal Ax_0}{g\mathcal Ax_0}\right) - \delta^{a3} \delta^{b3} \left(\cos g\mathcal Ax_0 - \frac{\sin g\mathcal Ax_0}{g\mathcal Ax_0}\right) \right. \\ \left. - \epsilon^{ab3} \left(\frac{\cos g\mathcal Ax_0-1}{g\mathcal Ax_0} + \sin g\mathcal Ax_0\right) \right) \varphi_1^b \\
	+ \frac1{g^2\mathcal Ax_0} \bigg( \delta^{ab} \sin g\mathcal Ax_0 + \delta^{a3} \delta^{b3} (g\mathcal Ax_0-\sin g\mathcal Ax_0) + \epsilon^{ab3} (\cos g\mathcal Ax_0-1) \bigg) \partial_\mu\varphi_1^b + \cdots \;.
\end{multline}
The dots contain higher-order terms in the quantum fields. In this expression, the coefficient of $\varphi_1^b$ turns out to be exactly the gradient of the coefficient of $\partial_\mu\varphi_1^b$, such that we can write more succinctly that
\begin{multline}
	(a^h)_\mu^a = \bigg( \delta^{ab} \cos g\mathcal Ax_0 - \epsilon^{ab3} \sin g\mathcal Ax_0 + \delta^{a3} \delta^{b3} (1-\cos g\mathcal Ax_0) \bigg) A_\mu^b \\
	+ \partial_\mu \left( \frac1{g^2\mathcal Ax_0} \bigg( \delta^{ab} \sin g\mathcal Ax_0 + \delta^{a3} \delta^{b3} (g\mathcal Ax_0-\sin g\mathcal Ax_0) + \epsilon^{ab3} (\cos g\mathcal Ax_0-1) \bigg) \varphi_1^b \right) + \cdots \;.
\end{multline}
In order to make this expression more transparent, let us introduce the following matrices:
\begin{subequations} \begin{gather}
	\mx R_3(g\mathcal Ax_0) = \begin{pmatrix} \cos g\mathcal Ax_0 & -\sin g\mathcal Ax_0 & 0 \\ \sin g\mathcal Ax_0 & \cos g\mathcal Ax_0 & 0 \\ 0 & 0 & 1 \end{pmatrix} \;, \label{rdef} \\
	\mx M = \frac1{g^2\mathcal Ax_0} \begin{pmatrix} \sin g\mathcal Ax_0 & \cos g\mathcal Ax_0-1 & 0 \\ -\cos g\mathcal Ax_0+1 & \sin g\mathcal Ax_0 & 0 \\ 0 & 0 & g\mathcal Ax_0 \end{pmatrix} \;.
\end{gather} \end{subequations}
With these definition, $a^h$ can be compactly written as
\begin{equation} \label{ahcompact}
	(a^h)_\mu^a = \mx R_3^{ab} (g\mathcal Ax_0) A_\mu^b + \partial_\mu (\mx M^{ab} \varphi_1^b) + \cdots \;.
\end{equation}

With the compact expression \eqref{ahcompact}, we can easily write down the gauge transform necessary to minimize the integral in \eqref{intasquared}. Minimizing \eqref{intasquared} under gauge transforms of $a_\mu^a$ is equivalent to demanding that $\partial_\mu (a^h)_\mu^a = 0$, or that
\begin{equation}
	\partial_\mu (\mx R_3^{ab} (g\mathcal Ax_0) A_\mu^b) + \partial^2 (\mx M^{ab} \varphi_1^b) = 0
\end{equation}
to this order in the quantum fields. We can formally solve for $\varphi_1^a$:
\begin{equation}
	\mx M^{ab} \varphi_1^b = - \frac1{\partial^2} \partial_\mu (\mx R_3^{ab} (g\mathcal Ax_0) A_\mu^b) \;,
\end{equation}
such that
\begin{align}
	(a^h)_\mu^a =& \mx R_3^{ab} (g\mathcal Ax_0) A_\mu^b - \partial_\mu \frac1{\partial^2} \partial_\nu (\mx R_3^{ab} (g\mathcal Ax_0) A_\nu^b) + \cdots \nonumber \\
	=& \left( \delta_{\mu\nu} - \frac{\partial_\mu\partial_\nu}{\partial^2} \right) (\mx R_3^{ab} (g\mathcal Ax_0) A_\nu^b) + \cdots \;. \label{ahfinal}
\end{align}
We thus see that $a^h$ is attained by first gauge transforming $A_\mu^a$ using the adjoint representation of the gauge transform that sets $\bar A_\mu^a$ equal to zero,\footnote{For the relevant background here considered, we thus have $(\bar A^h)_\mu^a=0$.} after which the result must be projected on its transversal space.

\subsection{Resulting action}
Let us now look at what the result \eqref{ahfinal} entails for the physics of the theory. After integrating out the localizing ghosts ($\phi$, $\bar\phi$, $\omega$, and $\bar\omega$) and the Nakanishi--Lautrup field $b$, we find that the part of the action quadratic in the quantum fields is given by
\begin{equation}\label{dv1}
	\int d^4x \left( A_\mu^a \left( - \delta_{\mu\nu} (\bar{\mathcal D}^2)^{ac} + \left(1-\frac1\xi\right) \bar{\mathcal D}_\mu^{ab}\bar{\mathcal D}_\nu^{bc} \right) A_\nu^c + \gamma^2 (a^h)_\mu^a \frac1{-\partial^2} (a^h)_\mu^a \right) \;.
\end{equation}
In order to get insight in the propagator, let us take the Fourier transform of the gluon field and use isospin eigenvectors. We again work in SU(2) with a background $\bar A_\mu^a = \mathcal A \delta^{a3} \delta_{\mu0}$. The first part of the quadratic action turns into the usual
\begin{multline}
	\int \frac{d^4p}{(2\pi)^4} \sum_{s=\pm,0} \tilde A_\mu^{-s}(-p) \left( - \delta_{\mu\nu} ((p_0-gs\mathcal A)^2 + \vec p^2) \phantom{\frac1\xi} \right. \\ \left. + \left(1-\frac1\xi\right) (p_\mu-gs\mathcal A\delta_{\mu0})(p_\nu-gs\mathcal A\delta_{\nu0}) \right) \tilde A_\nu^s(p) \;,
\end{multline}
where, with a slight abuse of notation, we mean $A_\mu^3$ for $A_\mu^s$ with $s$ equal to ``$0$''.

The second part, containing $a^h$, needs slightly more work. Using the definition \eqref{rdef} of the rotation matrix $\mx R_3$, we find
\begin{equation}
	(a^h)_\mu^s = \left(\delta_{\mu\nu} - \frac{\partial_\mu\partial_\nu}{\partial^2}\right) e^{isg\mathcal Ax_0} A_\nu^s \;.
\end{equation}
For the Fourier transformed function we find
\begin{multline}
	(\tilde a^h)_\mu^s(p_\lambda) = \int d^4x \; e^{ip_\mu x_\mu} \left(\delta_{\mu\nu} - \frac{\partial_\mu\partial_\nu}{\partial^2}\right) e^{isg\mathcal Ax_0} A_\nu^s(x) \\
	= \left(\delta_{\mu\nu} - \frac{p_\mu p_\nu}{p^2}\right) \int d^4x \; e^{ip_\mu x_\mu + isg\mathcal Ax_0} A_\nu^s(x) = \left(\delta_{\mu\nu} - \frac{p_\mu p_\nu}{p^2}\right) \tilde A_\nu^s(p_\lambda+sg\mathcal A\delta_{\lambda0}) \;,
\end{multline}
where we partially integrated the projector when going to the second line. The second contribution to the quadratic action is thus
\begin{equation}
	\gamma^2 \int \frac{d^4p}{(2\pi)^4} \sum_{s=\pm,0} \tilde A_\mu^{-s}(-p_\lambda-sg\mathcal A\delta_{\lambda0}) \left(\delta_{\mu\nu} - \frac{p_\mu p_\nu}{p^2}\right) \frac1{p^2} \tilde A_\nu^s(p_\lambda+sg\mathcal A\delta_{\lambda0}) \;.
\end{equation}
If we now perform a shift $p_\mu \to p_\mu-sg\mathcal A\delta_{\mu0}$ in the integration of the momenta, we finally find
\begin{equation} \label{finalgribovpart}
	\gamma^2 \int \frac{d^4p}{(2\pi)^4} \sum_{s=\pm,0} \tilde A_\mu^{-s}(-p) \left(\delta_{\mu\nu} - \frac{(p_\mu-sg\mathcal A\delta_{\mu0})(p_\nu-sg\mathcal A\delta_{\nu0})}{(p_0-sg\mathcal A)^2+\vec p^2}\right) \frac1{(p_0-sg\mathcal A)^2+\vec p^2} \tilde A_\nu^s(p) \;.
\end{equation}

We can conclude by stating that the part of the action quadratic in the quantum gauge field looks as if it were
\begin{equation} \label{finalall}
	\int d^4x \; A_\mu^a \left( - \delta_{\mu\nu} (\bar{\mathcal D}^2)^{ac} + \left(1-\frac1\xi\right) \bar{\mathcal D}_\mu^{ab}\bar{\mathcal D}_\nu^{bc} + \gamma^2 \left(\delta^{ab} \delta_{\mu\nu} - \left(\frac{\bar{\mathcal D}_\mu\bar{\mathcal D}_\nu}{\bar{\mathcal D}^2}\right)^{ab}\right) \left(\frac1{-\bar{\mathcal D}^2}\right)^{bc} \right) A_\nu^c \;.
\end{equation}
When computing the effective action of the theory at leading order, we will have to take the trace of the logarithm of the expression multiplying $A_\mu^a A_\nu^c$ in the previous expression. If we take the trace by going to Fourier space and integrating over the momenta, the presence of the background field $\bar A_\mu^a = \mathcal A \delta^{a3} \delta_{\mu0}$ will simply cause a shift in the timelike component of the momentum. At zero temperature, such a shift does not influence the final result, as it is simply a shift in an integration variable. This fact shows that background gauge invariance remains manifest, which is what we wanted.

It is perhaps interesting to stress here the difference with the usual practice of introducing a chemical potential via an \emph{imaginary}  temporal background field \cite{Kapusta:1981aa,Loewe:2005df}. In this case, the imaginary value cannot be shifted away anymore using the real-valued integration variable $p_0$. In gauge transformation language, it would correspond to using a U($1$) rotation angle that is imaginary, something which is evidently not allowed.

\section{Outlook}
Now that we have introduced a background gauge invariant Gribov--Zwanziger action in the Landau--DeWitt gauge, summarized in \eqref{symactionbis}, and have explicitly shown how to work with this action at zero temperature for at least the phenomenologically relevant constant Abelian backgrounds in the temporal direction, we can now progress in future work towards the finite temperature analysis, of relevance for probing  the deconfinement transition using functional tools. Doing so, we can correct the earlier analyses of \cite{Canfora:2015yia,Canfora:2016ngn} where the non-background gauge invariant Gribov--Zwanziger action was employed, with the here discussed spontaneous Lorentz symmetry breaking that is now averted.

Let us end by noticing that at non-zero temperature in the imaginary time (Matsubara) formalism, the background \eqref{dd0} can no longer be gauge transformed to zero and that the shift leading to \eqref{finalgribovpart} can no longer be done ---due to the periodicity constraints--- thereby complicating the computation of the effective action. Also the auxiliary bosonic and fermionic fields present in the action \eqref{symactionbis} will contribute in a non-trivial fashion. Another point of interest will be the addition of extra gauge (BRST) invariant $d=2$ condensates relevant for the dynamics \cite{Capri:2017bfd,Dudal:2008sp,Vercauteren:2010rk}. We will report on this elsewhere.

\section*{Acknowledgments}
We are grateful for inspiring discussions with F.~Canfora and P.~Pais. D.V.~is grateful for the hospitality at KU Leuven -- Kulak where this work was initiated.

\appendix

\section{Conventions}
We define isospin eigenstates as
\begin{equation} \label{isospin}
	\mx v_+ = \frac1{\sqrt2} \begin{pmatrix} 1 \\ i \\ 0 \end{pmatrix} \;, \qquad \mx v_- = \frac1{\sqrt2} \begin{pmatrix} 1 \\ -i \\ 0 \end{pmatrix} \;, \qquad \mx v_3 = \begin{pmatrix} 0 \\ 0 \\ 1 \end{pmatrix} \;.
\end{equation}
We then have that
\begin{subequations} \begin{gather}
	\mathbbm 1 = \mx v_+ \mx v_+^\dagger + \mx v_- \mx v_-^\dagger + \mx v_3 \mx v_3^\dagger \label{spinunity} \\
	\tr \mx A = \mx v_+^\dagger \mx A \mx v_+ + \mx v_-^\dagger \mx A \mx v_- + \mx v_3^\dagger \mx A \mx v_3 \;.
\end{gather} \end{subequations}
Furthermore, for the Levi--Civita tensor we have that
\begin{equation}
	\epsilon^{ab3} \mx v_3^b = 0 \;, \qquad \epsilon^{ab3} \mx v_s^b = is \mx v_s^a \;,
\end{equation}
and
\begin{equation}
	\epsilon^{ab3} A_\mu^b = i A_\mu^- \mx v_+^a - i A_\mu^+ \mx v_-^a \;,
\end{equation}
where
\begin{equation} \label{isospingluon}
	A_\mu^s = \frac1{\sqrt2} (A_\mu^1 + isA_\mu^2) \;.
\end{equation}

\bibliographystyle{unsrt}
\bibliography{Bibliografie}

\begin{thebibliography}{10}

\bibitem{Abbott:1980hw}
L.~F. Abbott.
\newblock {The Background Field Method beyond one loop}.
\newblock {\em Nucl. Phys.}, B185:189--203, 1981.

\bibitem{Abbott:1981ke}
L.~F. Abbott.
\newblock Introduction to the background field method.
\newblock {\em Acta Phys.Polon.}, B13:33, 1982.

\bibitem{Marhauser:2008fz}
F.~Marhauser and J.~M. Pawlowski.
\newblock {Confinement in Polyakov gauge}.
\newblock 2008.
\newblock arXiv:0812.1144 [hep-ph].

\bibitem{Braun:2007bx}
J.~Braun, H.~Gies, and J.~M. Pawlowski.
\newblock {Quark confinement from color confinement}.
\newblock {\em Phys. Lett.}, B684:262--267, 2010.

\bibitem{Reinosa:2014ooa}
U.~Reinosa, J.~Serreau, M.~Tissier, and N.~Wschebor.
\newblock {Deconfinement transition in SU($N$) theories from perturbation
  theory}.
\newblock {\em Phys. Lett.}, B742:61--68, 2015.

\bibitem{Reinosa:2015gxn}
U.~Reinosa, J.~Serreau, M.~Tissier, and N.~Wschebor.
\newblock {Two-loop study of the deconfinement transition in Yang--Mills
  theories: SU(3) and beyond}.
\newblock {\em Phys. Rev.}, D93(10):105002, 2016.

\bibitem{Reinhardt:2012qe}
H.~Reinhardt and J.~Heffner.
\newblock {The effective potential of the confinement order parameter in the
  Hamilton approach}.
\newblock {\em Phys. Lett.}, B718:672--677, 2012.

\bibitem{Reinhardt:2013iia}
H.~Reinhardt and J.~Heffner.
\newblock {Effective potential of the confinement order parameter in the
  Hamiltonian approach}.
\newblock {\em Phys. Rev.}, D88:045024, 2013.

\bibitem{Cyrol:2017qkll}
A.~K. Cyrol, M.~Mitter, J.~M. Pawlowski, and N.~Strodthoff.
\newblock {Non-perturbative finite-temperature Yang--Mills theory}.
\newblock 2017.
\newblock arXiv:1708.03482 [hep-ph].

\bibitem{Aguilar:2006gr}
A.~C. Aguilar and J.~Papavassiliou.
\newblock {Gluon mass generation in the PT-BFM scheme}.
\newblock {\em JHEP}, 12:012, 2006.

\bibitem{Aguilar:2008xm}
A.~C. Aguilar, D.~Binosi, and J.~Papavassiliou.
\newblock {Gluon and ghost propagators in the Landau gauge: Deriving lattice
  results from Schwinger-Dyson equations}.
\newblock {\em Phys. Rev.}, D78:025010, 2008.

\bibitem{Binosi:2009qm}
D.~Binosi and J.~Papavassiliou.
\newblock {Pinch technique: Theory and applications}.
\newblock {\em Phys. Rept.}, 479:1--152, 2009.

\bibitem{brstbackground}
P.~A. {Grassi}, T.~{Hurth}, and A.~{Quadri}.
\newblock {Landau background gauge fixing and the IR properties of Yang--Mills
  Green functions}.
\newblock {\em Phys.Rev.}, D70:105014, 2004.

\bibitem{Ferrari:2000yp}
R.~Ferrari, M.~Picariello, and A.~Quadri.
\newblock {Algebraic aspects of the background field method}.
\newblock {\em Annals Phys.}, 294:165--181, 2001.

\bibitem{Binosi:2012st}
D.~Binosi and A.~Quadri.
\newblock {The Background Field Method as a Canonical Transformation}.
\newblock {\em Phys. Rev.}, D85:121702, 2012.

\bibitem{Gribov:1977wm}
V.~N. Gribov.
\newblock {Quantization of nonabelian gauge theories}.
\newblock {\em Nucl.Phys.}, B139:1, 1978.

\bibitem{Zwanziger:1989mf}
D.~Zwanziger.
\newblock {Local and renormalizable action from the Gribov horizon}.
\newblock {\em Nucl.Phys.}, B323:513--544, 1989.

\bibitem{Zwanziger:1992qr}
D.~Zwanziger.
\newblock {Renormalizability of the critical limit of lattice gauge theory by
  BRS invariance}.
\newblock {\em Nucl.Phys.}, B399:477--513, 1993.

\bibitem{Vandersickel:2012tz}
N.~Vandersickel and D.~Zwanziger.
\newblock {The Gribov problem and QCD dynamics}.
\newblock {\em Phys. Rept.}, 520:175--251, 2012.

\bibitem{Zwanziger:1982na}
D.~Zwanziger.
\newblock {Nonperturbative modification of the Faddeev--Popov formula and
  banishment of the naive vacuum}.
\newblock {\em Nucl. Phys.}, B209:336, 1982.

\bibitem{Canfora:2015yia}
F.~E. Canfora, D.~Dudal, I.~F. Justo, P.~Pais, L.~Rosa, and D.~Vercauteren.
\newblock {Effect of the Gribov horizon on the Polyakov loop and vice versa}.
\newblock {\em Eur. Phys. J.}, C75(7):326, 2015.

\bibitem{Canfora:2016ngn}
F.~E. Canfora, D.~Hidalgo, and P.~Pais.
\newblock {The Gribov problem in presence of background field for $SU(2)$
  Yang--Mills theory}.
\newblock {\em Phys. Lett.}, B763:94--101, 2016.

\bibitem{Capri:2012wx}
M.~A.~L. Capri, D.~Dudal, M.~S. Guimar\~aes, L.~F. Palhares, and S.~P. Sorella.
\newblock {An all-order proof of the equivalence between Gribov's no-pole and
  Zwanziger's horizon conditions}.
\newblock {\em Phys.Lett.}, B719:448--453, 2013.

\bibitem{Capri:2015ixa}
M.~A.~L. Capri, D.~Dudal, D.~Fiorentini, M.~S. Guimar\~aes, I.~F. Justo, A.~D.
  Pereira, B.~W. Mintz, L.~F. Palhares, R.~F. Sobreiro, and S.~P. Sorella.
\newblock {Exact nilpotent nonperturbative BRST symmetry for the
  Gribov--Zwanziger action in the linear covariant gauge}.
\newblock {\em Phys. Rev.}, D92(4):045039, 2015.

\bibitem{Capri:2016aqq}
M.~A.~L. Capri, D.~Dudal, D.~Fiorentini, M.~S. Guimar\~aes, I.~F. Justo, A.~D.
  Pereira, B.~W. Mintz, L.~F. Palhares, R.~F. Sobreiro, and S.~P. Sorella.
\newblock {Local and BRST-invariant Yang-Mills theory within the Gribov
  horizon}.
\newblock {\em Phys. Rev.}, D94(2):025035, 2016.

\bibitem{Capri:2017bfd}
M.~A.~L. Capri, D.~Fiorentini, A.~D. Pereira, and S.~P. Sorella.
\newblock {Renormalizability of the refined Gribov--Zwanziger action in linear
  covariant gauges}.
\newblock {\em Phys. Rev.}, D96(5):054022, 2017.

\bibitem{Lavelle:1995ty}
M.~Lavelle and D.~McMullan.
\newblock {Constituent quarks from QCD}.
\newblock {\em Phys. Rept.}, 279:1--65, 1997.

\bibitem{vanBaal:1991zw}
P.~van Baal.
\newblock {More (thoughts on) Gribov copies}.
\newblock {\em Nucl. Phys.}, B369:259--275, 1992.

\bibitem{Kapusta:1981aa}
J.~I. Kapusta.
\newblock {Bose--Einstein condensation, spontaneous symmetry breaking, and
  gauge theories}.
\newblock {\em Phys. Rev.}, D24:426--439, 1981.

\bibitem{Loewe:2005df}
M.~Loewe, S.~Mendizabal, and J.~C. Rojas.
\newblock {Background field method at finite temperature and density}.
\newblock {\em Phys. Lett.}, B635:213--217, 2006.

\bibitem{Dudal:2008sp}
D.~Dudal, J.~A. Gracey, S.~P. Sorella, N.~Vandersickel, and H.~Verschelde.
\newblock {A refinement of the Gribov--Zwanziger approach in the Landau gauge:
  Infrared propagators in harmony with the lattice results}.
\newblock {\em Phys.Rev.}, D78:065047, 2008.

\bibitem{Vercauteren:2010rk}
D.~Vercauteren and H.~Verschelde.
\newblock {The asymmetry of the dimension 2 gluon condensate: The finite
  temperature case}.
\newblock {\em Phys. Rev.}, D82:085026, 2010.

\end{thebibliography}

\end{document}